# Two-dimensional dispersive shock waves in dissipative optical media

Yaroslav V. Kartashov[1,2,*] and Anatoly M. Kamchatnov[2]

[1]ICFO-Institut de Ciencies Fotoniques, and Universitat Politecnica de Catalunya, Mediterranean Technology Park, 08860 Castelldefels (Barcelona), Spain
[2]Institute of Spectroscopy, Russian Academy of Sciences, Troitsk, Moscow Region, 142190, Russia

We study generation of two-dimensional dispersive shock waves and oblique dark solitons upon interaction of tilted plane waves with negative refractive index defects embedded into defocusing material with linear gain and two-photon absorption. Different evolution regimes are encountered including the formation of well-localized disturbances for input tilts below critical one, and generation of extended shock waves containing multiple intensity oscillations in the "upstream" region and gradually vanishing oblique dark solitons in "downstream" region for input tilts exceeding critical one. The generation of stable dispersive shock waves is possible only below certain critical defect strength.
  OCIS Codes: 190.5940, 190.6135

Dispersive shock waves may be encountered in physical settings where nonlinearity may lead to wave breaking of a pulse for certain input conditions. Around such wave breaking points the dispersive effects become particularly strong and lead to regularization of singularity. This process is accompanied by the appearance of extended oscillating patterns connecting two smooth distributions with different parameters; these patterns are called as dispersive shock waves. In optics shock waves were theoretically analyzed and experimentally observed in various settings, in both temporal [1-3] and spatial [4-7] domains. In particular, for observation of spatial shock waves photorefractive [5] and nonlocal [6,7] materials were utilized. Besides that, this problem was under active investigation in physics of Bose-Einstein condensates [8-13], where dispersive shock waves were observed too.

Interaction of waves with two-dimensional defects is accompanied by many interesting effects, which cannot be observed in one-dimensional case. Depending on the velocity of the flow and defect strength, one may observe formation of localized disturbances, generation of vortices, ship waves, and oblique dark solitons, as it was predicted for matter [10,11] and optical [14] waves. Most of previous studies of shock waves generated on two-dimensional defects were concentrated on conservative systems. However, recent experiments on the interaction of dissipative polariton condensates with inhomogeneities in the cavity have revealed important differences in excitation dynamics of shock waves in dissipative media in comparison with conservative one [15,16].

In this Letter we, using a two-dimensional generalization of simple model studied in [17,18], show that a stationary two-dimensional dispersive shock waves may be generated upon interaction of tilted plane waves with negative refractive index defects in the material with gain and two-photon absorption. We use the nonlinear Schrödinger equation for the dimensionless light field amplitude $q$ for the description of light propagation along the $\xi$ axis in defocusing cubic medium with gain, two-photon absorption, and negative refractive index defect:

$$i\frac{\partial q}{\partial \xi} = -\frac{1}{2}\left(\frac{\partial^2 q}{\partial \eta^2} + \frac{\partial^2 q}{\partial \zeta^2}\right) + (\sigma_r - i\sigma_i)q|q|^2 + [ip_i + p_r R(\eta,\zeta)]q, \quad (1)$$

where $\eta, \zeta$ are the transverse coordinates scaled to characteristic width $r_0$, $\xi$ is the propagation distance normalized to the diffraction length $2\pi n r_0^2/\lambda$, $\sigma_r = 1$ corresponds to defocusing nonlinearity, $\sigma_i$ is the strength of two-photon absorption, $p_i$ is the gain parameter, $p_r$ characterizes the strength of refractive index defect, while the function $R(\eta,\zeta)$ describes the profile of the defect. Here we consider broad and smooth defects described by $R(\eta,\zeta) = \exp[-(\eta^2+\zeta^2)/d^2]$ with $d=5$ and set $\sigma_i = p_i = 0.05$. For waves at $\lambda = 0.61\,\mu\text{m}$ and $r_0 = 10\,\mu\text{m}$ propagating in CdS samples with $n_2 \approx -10^{-17}\,\text{m}^2/\text{W}$, the distance $\xi=1$ corresponds to 2.6 mm of propagation, $q \sim 1$ corresponds to peak intensity $\sim 3 \times 10^{12}\,\text{W/m}^2$, $p_i = 0.05$ corresponds to gain coefficient $\sim 38\,\text{m}^{-1}$ and $\sigma_i = 0.05$ corresponds to absorption $\sim 1.2 \times 10^{-11}\,\text{m/W}$.

In the absence of refractive index defect the Eq. (1) admits stationary solution in the form of tilted plane wave $q(\eta,\zeta,\xi) = \chi \exp(i\alpha\eta + i\beta\zeta + ib\xi)$, with amplitude $\chi = (p_i/\sigma_i)^{1/2}$ and propagation constant $b = -\chi^2 - (\alpha^2 + \beta^2)/2$. This solution is stable as can be checked by substitution of perturbed wave $q = [\chi + u\exp(i\mathbf{k}\mathbf{r} + \delta\xi) + v^* \exp(-i\mathbf{k}\mathbf{r} + \delta^*\xi)]\exp(i\alpha\eta + i\beta\zeta + ib\xi)$ with $\mathbf{k} = (k_\eta, k_\zeta)$ and $\mathbf{r} = (\eta,\zeta)$ into Eq. (1), its linearization and solution of resulting linear eigenvalue problem, which for particular case $\alpha, \beta = 0$ gives the following dispersion relation:

$$\delta = -p_i \pm [p_i^2 - (|\mathbf{k}|^2/4)(|\mathbf{k}|^2 + 4p_i/\sigma_i)]^{1/2} \quad (2)$$

As one can see, $\text{Re}\,\delta \leq 0$ that indicates on stability. This dispersion relation allows to introduce the critical tilt as a positive value of the limit $\alpha_{cr} = \lim_{|\mathbf{k}|\to 0} \lim_{p_i,\sigma_i \to 0}(i\delta/|\mathbf{k}|)$ that corresponds to nearly linear dependence on $\mathbf{k}$ in the dispersion relation $\delta(|\mathbf{k}|)$ at $p_i, \sigma_i \ll 1$, $p_i/\sigma_i = \text{const}$. In our case the critical tilt is determined by the plane wave amplitude $\alpha_{cr} = \chi = 1$. In the presence of refractive index defect the evolution of input wave $q|_{\xi=0} = \chi\exp(i\alpha\eta)$ is qualitatively different for $\alpha < \alpha_{cr}$ and $\alpha > \alpha_{cr}$.

Here we investigate the formation of stationary two-dimensional shock waves upon interaction of tilted plane waves with negative refractive index defects. Since the system is dissipative, it possesses attractors in certain parameter region and the input wave usually evolves toward such an attractor upon propagation. In order to obtain the entire family of shock wave solutions we slowly increased defect strength $p_r$ and for each $p_r$ solved Eq. (1) directly using split-step fast Fourier method up to huge distance $\xi \sim 10^3 - 10^4$ until the formation of stationary wave was observed. To speed up formation of stationary structures we used the output wave from previous step in $p_r$ as an input for new defect strength $p_r + \delta p_r$.

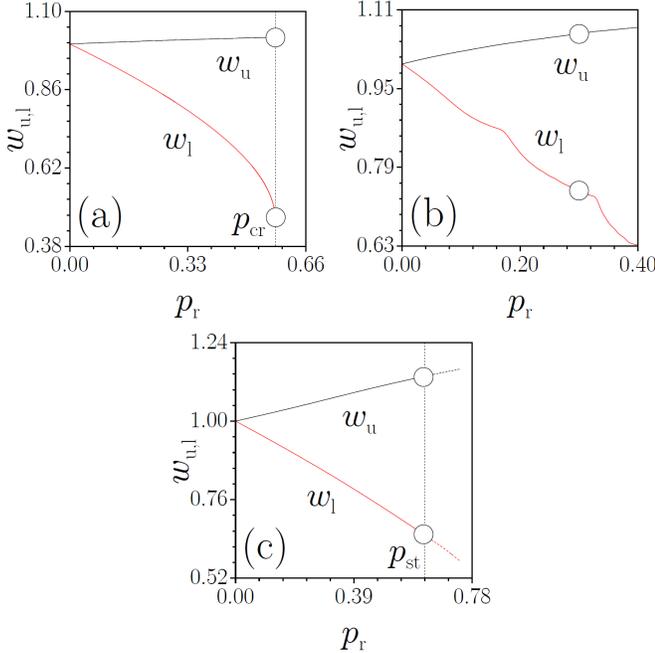

Fig. 1. Maximal $w_u$ and minimal $w_l$ amplitudes of shock wave versus defect depth at $\alpha = 0.5$ (a), $\alpha = 1.0$ (b), and $\alpha = 1.5$ (c). Vertical dashed lines in (a) and (c) mark the border of existence and stability domains for such waves, respectively. Circles correspond to waves shown in Figs. 2(a)-2(c).

One can characterize stationary waves with the dependencies of maximal $w_u = \max|q|$ and minimal $w_l = \min|q|$ amplitudes on $p_r$. These dependencies for different regimes $\alpha < \alpha_{cr}$, $\alpha = \alpha_{cr}$, and $\alpha > \alpha_{cr}$ are shown in Fig. 1, while Fig. 2 illustrates typical shapes of stationary two-dimensional waves. In subcritical regime $\alpha < \alpha_{cr}$ the wave contains intensity deep centered slightly after the defect and having characteristic size of the order of defect width, and a hump in the "upstream" region (Fig. 2, top left). Since the system is dissipative the wave is strongly perturbed around the defect, while far from it the amplitude approaches the asymptotic value $\chi = (p_l / \sigma_i)^{1/2}$ dictated by the balance of gain and losses. The deep becomes more and more pronounced with increase of $p_r$ (in some cases the minimal amplitude may decrease even up to $w_l \sim 0.2\chi$), while the maximal amplitude of the wave only slightly grows [Fig. 1(a)]. At $p_r = p_r^{cr}$ [dashed line in Fig. 1(a)] the family of stationary waves ceases to exist and the tangential line to $w_l(p_r)$ dependence becomes vertical. Thus, there exists critical defect depth above which stationary two-dimensional waves were not found for subcritical tilts. The critical defect depth is the monotonically decreasing function of tilt [Fig. 3(a)], i.e. the range of existence of stationary disturbances shrinks when $\alpha \to \alpha_{cr}$. This behavior was encountered for $0.3 \leq \alpha \leq 0.9$. Within this range, the transition to strong defect potentials with $p_r > p_r^{cr}$ leads to non-stationary regimes with periodic generation of vortex pairs. However, for $\alpha < 0.3$ the cutoff on $p_r$ does not exist and stationary waves exist even for $p_r \gg 1$.

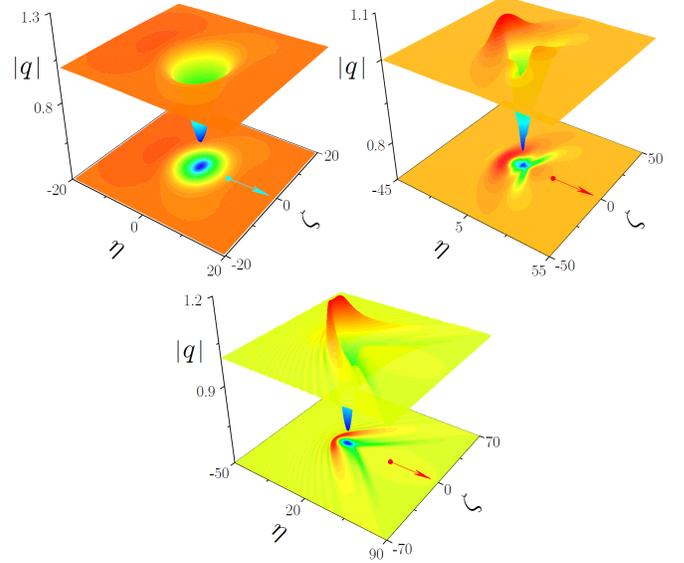

Fig. 2. Field modulus distributions in stationary wave structures emerging due to the presence of defect at $\alpha = 0.5$, $p_r = 0.575$ (top left panel), (b) $\alpha = 1.0$, $p_r = 0.3$ (top right panel), and (c) $\alpha = 1.5$, $p_r = 0.62$ (bottom panel). Arrows on the projection plane indicate direction of the energy flow.

At $0.9 < \alpha < 1.1$ and, in particular, for critical angle $\alpha = \alpha_{cr}$ the dependencies $w_{u,l}(p_r)$ change qualitatively [Fig. 1(b)]. Instead of appearance of cutoff, one observes oscillations in the minimal amplitude of stationary wave and the dependencies $w_{u,l}$ can be continued to relatively large defect strengths. The oscillations in $w_l(p_r)$ are accompanied by the appearance of several local minima in the shape of the wave (Fig. 2, top right), whose number increases with increase of the defect strength. At critical tilt the amplitude of the hump in the "upstream" region is much larger than in subcritical regime and one can observe the tendency for formation of oblique dark solitons (two oblique stripes in the "downstream" region, which gradually disappear at $\eta \to +\infty$).

The formation of extended two-dimensional shock waves is observed for sufficiently large supercritical tilts (see an example in Fig 2, bottom). Such waves are characterized by multiple intensity oscillations (usually called ship waves) in the "upstream" region and the presence of oblique dark solitons in the "downstream" region. In most of stable stationary patterns that we obtained there were two clearly observable symmetric dark solitons. The transverse decay length (recall that at

$r \to \infty$ the amplitude $|q| \to \chi$) for both ship waves and dark solitons considerably exceeds the width of the refractive index defect. The length of dark solitons and the amplitude of ship waves increase with increase of defect strength. The total amplitude of two-dimensional shock wave also grows with $p_r$ [Fig. 1(c)]. For fixed $p_r$ the amplitude $w_u - w_l$ of shock wave decreases with increase of tilt $\alpha$ and for $\alpha \sim 3\alpha_{cr}$ the main intensity hump in the "upstream" region becomes nearly comparable in amplitude with intensity deep just after the defect. We found that in supercritical regime the instability develops if the defect strength exceeds the value $p_r = p_r^{st}$ [the border of stability domain is indicated by dashed line in Fig. 1(c)], although shock wave family exists even at $p_r > p_r^{st}$ in contrast to subcritical regime where one finds cutoff $p_r^{cr}$. The increase of the input tilt results in expansion of stability domain [see Fig. 3(b) showing the upper border of stability domain on the plane $(\alpha, p_r)$].

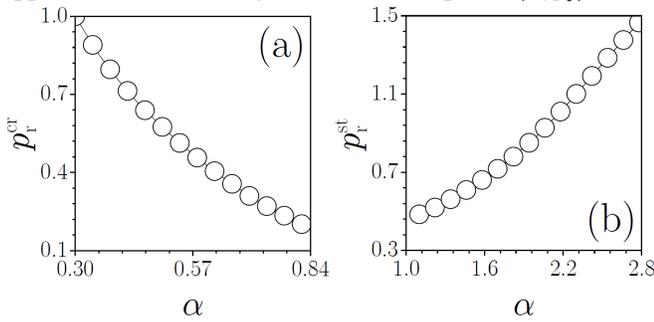

Fig. 3. (a) Critical value of defect depth $p_r^{cr}$ at which stationary disturbances cease to exist for velocities $\alpha < 1$. (b) The defect depth $p_r^{st}$ at which stationary shock waves become unstable for velocities $\alpha > 1$.

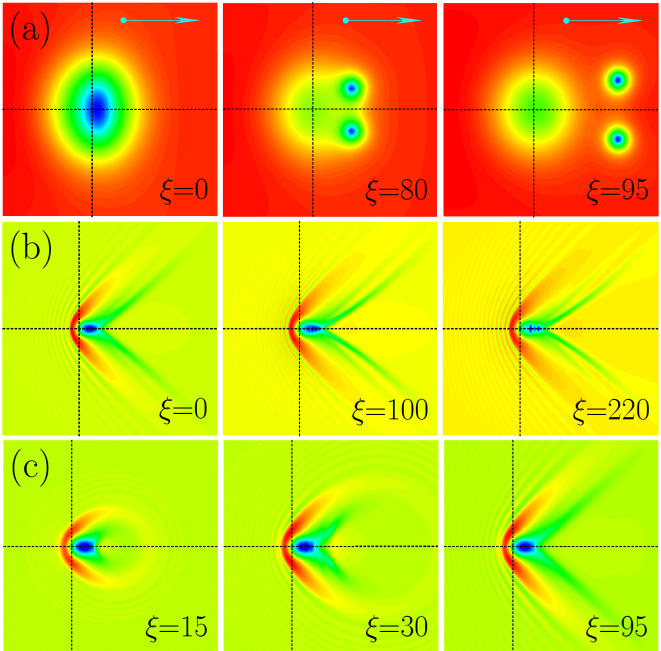

Fig. 4. Unstable evolution of the wave at $\alpha = 0.5$, $p_r = 0.58$ (a) and $\alpha = 1.5$, $p_r = 0.90$ (b). In (a) the input condition corresponds to the disturbance that propagates stably at $\alpha = 0.5$, $p_r = 0.57$, while in (b) the input condition corresponds to shock wave that is stable at $\alpha = 1.5$, $p_r = 0.62$. (c) The excitation of shock wave by tilted plane wave at $\alpha = 1.5$, $p_r = 0.45$. The point where dashed lines cross indicate the center of the defect. Arrows in top row show the direction of the energy flow.

Various evolution scenarios possible in our setting are depicted in Fig. 4. If the strength of the defect exceeds $p_r^{cr}$ for subcritical tilts $\alpha < \alpha_{cr}$ one usually observes periodic emission of vortex pairs [Fig. 4(a)]. For supercritical tilts $\alpha > \alpha_{cr}$ and defect strengths $p_r > p_r^{st}$ lying inside instability domain one initially observes nearly stationary pattern where more than one oblique dark solitons are resolved [Fig. 4(b)]. Finally, the formation of truly stationary stable shock wave from tilted plane wave in proper parameter region is depicted in Fig. 4(c). It is worth noticing that above observations have clear hydrodynamic counterparts which can be applied to dynamics of microcavity polaritons and other quantum fluids. In terms of this physics, $\alpha$ has a meaning of Mach number of incoming flow and $p_r$ measures the strength of the obstacle potential. With increase of the Mach number and obstacle strength the transition occurs from a laminar stationary flow to a nonstationary shedding of vortices.

Summarizing, we showed that stable dispersive shock waves may form upon interaction of tilted plane waves with refractive index defects in weakly dissipative media. In both subcritical and supercritical regimes there exists maximal defect strength below which two-dimensional waves propagate stably.